\documentclass[pra,twocolumn,showpacs,showkeys,floatfix,amsmath,amssymb]{revtex4}
\usepackage{graphicx,exscale,bm,color}
\begin{document}
%-----------------------------------------------------------------------------
\title{Time-optimal bath-induced unitaries by Zermelo
       navigation: \\speed limit and non-Markovian quantum computation}
\date{\today}
\author{Jens Clausen}
\affiliation{Institut f\"ur Theoretische Physik,
             Universit\"at Innsbruck,
             Technikerstra\ss{}e 21a,
             A-6020 Innsbruck,
	     Austria}
%\author{Jens Clausen, Durga Dasari Bhaktavatsala Rao, and Gershon Kurizki}
%\affiliation{Department of Chemical Physics,
%             Weizmann Institute of Science,
%             Rehovot,
%             76100, Israel}
\begin{abstract}
The solution of the quantum Zermelo navigation problem is applied to the
non-Markovian open system dynamics of a set of quantum systems interacting with
a common environment. We consider a case allowing an exact time-optimal
realization of environment-mediated non-local system unitaries. For a linear
coupling to a harmonic bosonic bath, we derive a speed limit for the
implementation time in terms of the fundamental frequency of the bath modes.
As a product of two exponentials of the local free wind and the pairwise
system-coupling, the Zermelo unitary forms a natural building block for reaching
a general unitary by concatenation.
\end{abstract}
\maketitle
%\tableofcontents
%-----------------------------------------------------------------------------
\section{
\label{sec1}
Introduction}
%-----------------------------------------------------------------------------
The discussion of optimal control protocols for evolving quantum systems and the
relevance of transition speeds and their limits can be traced back to the dawn
of quantum theory as can be seen in the example of Fermi's golden rule for the
mean transition rate into an orthogonal energy eigenstate. While touching
fundamental problems about computation \emph{per se} \cite{Llo00,Mar14,Hor14},
it is also of technological interest, since minimization of computing time
amounts to minimizing the interaction time with detrimental environmental noise
sources.
The speed optimization can refer to the transformation of a given initial to a
target state, in particular the evolution into an (arbitrary) orthogonal state,
or it can refer to the implementation of a desired target unitary, which is
supposed to act on an arbitrary (unknown) quantum state.
The existence of a physical solution to the optimization problem and the
formulation of finite speed limits generally require a set of preset constraints
which represent practical limitations encountered. They can be of energetic
nature restricting the size of the Hamiltonians in some way or refer to the type
(subspace) of Hamiltonians that one can implement. The latter kind of
constraints can be added by Lagrange multipliers \cite{Wan14,Bro14a,Rus14b}
giving rise to quantum brachistochrone curves as optimal trajectories of states
or unitaries.

Besides of this, a fundamental constraint addresses external background fields
that cannot be manipulated and give rise to a free (natural) Hamiltonian
$\hat{H}_0$, which must be taken as given.
Even in the absence of restrictions on the implementable control Hamiltonians
$\hat{H}_{\mathrm{C}}(t)$ (which would suggest to simply substitute
$\hat{H}_{\mathrm{C}}^\prime$ $\!=$ $\!\hat{H}_{\mathrm{C}}$ $\!-$
$\!\hat{H}_0$), the mentioned energetic constraints limit the size of
implementable control fields, hence the generic $\hat{H}_0$ cannot be ignored.
This is the motivation for the quantum Zermelo navigation problem which,
inspired by a result in \cite{She03}, was formulated in \cite{Rus14a} and
recently solved in \cite{Rus14b,Bro14a}.

Roughly speaking, external forces as caused by a background field are
incorporated into the problem description as a geometric property by defining an
appropriate measure of distance with respect to which one then freely takes a
shortest path. In the example of the Zermelo navigation problem, the solution
consists in following the geodesics of a Randers metric, which is a special type
of Finsler metric and can be derived from Riemann metrics by adding a linear
term. We refer to \cite{Rus14b,Bro14a} and references therein for a description
of the mathematical background.

We here consider the navigation of unitary operators. Starting from a given
initial unitary $\hat{U}_{\mathrm{I}}$, the task is to find a time-dependent
Hamiltonian $\hat{H}(t)$ that implements a desired final unitary
$\hat{U}_{\mathrm{F}}$ in shortest possible time. The total Hamiltonian
\begin{equation}
\label{ZH}
  \hat{H}(t)=\hat{H}_0+\hat{H}_{\mathrm{C}}(t)
\end{equation}
is assumed to consist of a constant part $\hat{H}_0$ that cannot be altered and
a time-dependent part $\hat{H}_{\mathrm{C}}(t)$ which can be controlled by the
experimenter without limits except that the size of the control is bounded, so
that we set a fixed $\mathrm{Tr}[\hat{H}_{\mathrm{C}}^2(t)]$.
%and assume that
%$\mathrm{Tr}[\hat{H}_{\mathrm{C}}^2(t)]$ $\!>$ $\!\mathrm{Tr}[\hat{H}_0^2]$.

The reference to Zermelo originates in the classical problem of finding a
time-dependent heading of a ship or aircraft that, starting from some initial
location, navigates it to a destination in shortest possible time \cite{Zer31}.
$\hat{H}_0$ here corresponds to a wind or current, while a constant
$\mathrm{Tr}[\hat{H}_{\mathrm{C}}^2(t)]$ corresponds to the assumption of
``full speed ahead'' at any time. In the classical problem, the optimal route
under constant wind or current is a straight line from start to finish, which is
achieved by keeping a constant correction angle of the vessel's heading with
respect to its destination to compensate the drift off. More general versions of
the classical Zermelo navigation problem that include obstacles and a wind
depending on location and time have also been of recent interest in animal
behavior research \cite{Hay14} and computer science \cite{Li13}.

Contrary to intuition, the solution of the corresponding quantum problem is
\emph{not} a constant
$\hat{H}_{\mathrm{C}}(t)$ $\!\equiv$ $\!\hat{H}_{\mathrm{C}}$ which
would give rise to a (single-exponential) one-parameter subgroup of unitaries
$\hat{U}(t)$ $\!=$
$\!\mathrm{T}\mathrm{e}^{-\mathrm{i}\int_{0}^{t}\!\mathrm{d}\tau\hat{H}(\tau)}$
$\!=$ $\!\mathrm{e}^{-\mathrm{i}(\hat{H}_0+\hat{H}_{\mathrm{C}})t}$, but instead
an explicit time-dependence such that $\hat{H}$ becomes time-independent
in the interaction picture
\cite{Rus14b,Bro14a},
\begin{equation}
\label{Zermelo}
  \hat{H}_{\mathrm{C}}(t)=\mathrm{e}^{-\mathrm{i}\hat{H}_0t}
  \hat{H}_{\mathrm{C}}(0)
  \mathrm{e}^{\mathrm{i}\hat{H}_0t},
\end{equation}
%given by the solution of the quantum Zermelo-Euler-Poincar\'{e} equation
%$\dot{\hat{H}}_{\mathrm{C}}(t)$ $\!=$
%$\!-\mathrm{i}[\hat{H}_0,\hat{H}_{\mathrm{C}}(t)]$ ($\hbar$ $\!=$ $\!1$)
and for which the time-optimal curve of unitaries disentangles according to
$\hat{U}(t)$ $\!=$ $\!\mathrm{T}\mathrm{e}^{-\mathrm{i}
\int_{0}^{t}\!\mathrm{d}\tau\hat{H}(\tau)}$ $\!=$
$\!\mathrm{e}^{-\mathrm{i}\hat{H}_0t}
\mathrm{e}^{-\mathrm{i}\hat{H}_{\mathrm{C}}(0)t}$ $\!=$
$\!\mathrm{e}^{-\mathrm{i}\hat{H}_{\mathrm{C}}(t)t}
\mathrm{e}^{-\mathrm{i}\hat{H}_0t}$ into two exponentials.
(Here, T denotes positive time ordering, and we have set $\hbar$ $\!=$ $\!1$.)
Time-independence as in the classical case is thus only recovered for 
$[\hat{H}_0,\hat{H}_{\mathrm{C}}(0)]$ $\!=$ $\!0$.

Zermelo navigation of quantum \emph{states} has been solved in \cite{Bro14c}
also deriving a limit on the passage time, by tracing it back to the navigation
of unitaries. The energetic constraint here takes the form of a fixed variance
$\langle\Delta\hat{H}_{\mathrm{C}}^2\rangle$. The special case of
time-independent $\hat{H}_{\mathrm{C}}$ has been discussed in \cite{Bro14b},
placing focus on two-level systems and demonstrating that the treatment is
astonishingly nontrivial even with these restrictions. Interestingly, Eq.(5) in
\cite{Bro14c} transforms the state navigation problem
$|\Psi_{\mathrm{F}}\rangle$ $\!=$ $\!\hat{U}(t)|\Psi_{\mathrm{I}}\rangle$
(in its general form) into that of hitting a (counter)moving target,
$\mathrm{e}^{-\mathrm{i}\hat{H}_{\mathrm{C}}(0)t}|\Psi_{\mathrm{I}}\rangle$
$\!=$ $\!\mathrm{e}^{\mathrm{i}\hat{H}_0t}|\Psi_{\mathrm{F}}\rangle$, which
suggests a close connection to pursuit problems: it advises to move
``straightly'' to where the target will be at the time of arrival. This is
analogous to the classical case of a dog who should straightly move towards the
extrapolated target (her walking master's) location, rather than intuitively
following her snout's direction heading towards the instantaneous location of
the target resulting in a longer so-called ``Hundekurve''.

Referring to time-independent systems with Hamiltonian $\hat{H}$, \cite{Rus14c}
relates speed limits for the implementation of unitaries to a family of bounds
on orthogonality times, which are all of the canonical form $t_{\perp}$ $\!\ge$
$\!\frac{\pi\hbar}{2E}$, including the Margolus-Levitin theorem \cite{Mar98}
({\small $E$ $\!=$ $\!\langle\psi|\hat{H}|\psi\rangle$ $\!-$ $\!E_0$, $E_0$}
being the lowest eigenvalue of $\hat{H}$), while relation to the time-energy
uncertainty relation (Mandelstam-Tamm bound \cite{Man45}, {\small $E$ $\!=$
$\!\sqrt{\langle\psi|\hat{H}^2|\psi\rangle-\langle\psi|\hat{H}|\psi\rangle^2}$})
is also discussed.

Here, we show how the Zermelo navigation (\ref{Zermelo}) can be applied to a
case of non-Markovian evolution of open quantum systems \cite{bookBreuer} such
that it allows an exact treatment of a set of quantum systems coupled to a
common environment. In contrast to the original motivation for realizing a
time-optimal evolution, namely to minimize the temporal accumulation of
environment-induced decoherence effects, we demonstrate that Zermelo
time-optimality can enable a \emph{deliberate} environmental coupling, which
here serves to mediate a desired system coupling. This intended utilization of
non-Markovian (memory) properties of the environment thus contrasts the case,
where environmental interactions are accompanied with irreversible decoherence.
Bath-induced system interactions have been discussed for several years,
cf. e.g. \cite{Bri01,Bra02}. The main focus of these Markovian approaches has
been on the preparation of entangled many-body steady states
\cite{Die08,Kra08,Kra11,Mus11} for dissipative quantum computation \cite{Ver09}
and continuous quantum repeaters \cite{Vol11}.

In contrast, here we consider a navigation of unitaries acting on a set of
systems and a shared bath. Starting from the identity,
$\hat{U}_{\mathrm{I}}$ $\!=$ $\!\hat{I}$, the navigation periodically passes
unitaries $\hat{U}(t_m)$ in which system and bath parts are refactorized.
Although the systems never interact directly, the system part of the target
unitary contains an environment-induced pairwise system-coupling which is
decoherence-free. All results are independent of the system and bath states and
apply exact analytical expressions which do not rely on assumptions such as the
weak coupling perturbative approach, rotating wave or Markovian approximations.

The present work builds on \cite{Dur11a,Nir11,Dur11b}, who demonstrate the
preparation of nonclassical states such as the GHZ-state, cf. also \cite{Zho05}.
However, the scheme presented there requires setting $\hat{H}_0$ $\!=$ $\!0$ to
allow for an analytic solution. Below we will show that (a) it is remarkably the
Zermelo navigation (\ref{Zermelo}) that enables us to overcome this limitation,
allowing for arbitrary $\hat{H}_0$. As a consequence, the described protocol of
generating bath-mediated system-coupling is time-optimal, a result thus also
holding for the schemes in \cite{Dur11a,Nir11,Dur11b} as a special case of
vanishing $\hat{H}_0$. Based on this, we construct (b) a
\emph{state-independent} speed limit for the refactorization time.
Finally, we address the question of reaching an arbitrarily given target
$\hat{U}_{\mathrm{F}}$. While a single Zermelo navigation $\hat{U}(t)$ allows
in our case merely the creation of bath-induced pairwise unitary system
interactions, the fact that $\hat{U}(t)$ is a product of two non-commuting
exponentials suggests the reachability of a given $\hat{U}_{\mathrm{F}}$ by
concatenation as realized by Hamiltonian resets. The two-exponential form of
$\hat{U}(t)$ thus halves the number of required resets. We will (c) provide a
numerical proof of principle demonstration, where the systems are three
two-level systems (``qubits''), and $\hat{U}_{\mathrm{F}}$ is a Toffoli gate.
Apart from being optimal for quantum error correction, the Toffoli gate alone is
universal for reversible computing and, together with the Hadamard gate, forms a
universal set of gates for quantum computing \cite{Zah15}. While the three
qubit-setup here serves as simplest case going beyond pairwise system
interactions, it is hence already sufficient as building block for universal
quantum computation.

This work is organized as follows. After this introduction, the realization of
an exact decoherence-free bath-mediated unitary system coupling by means of
Zermelo navigation is presented in Sec.~\ref{sec2}. Sec.~\ref{sec2.1} defines
the general setting and conditions, Sec.~\ref{sec2.2} introduces the example of
a linear coupling to a bosonic bath, for which the speed limit is derived in
Sec.~\ref{sec2.3}. A numerical approach to the problem of navigating to a
general target is given in Sec.~\ref{sec3}. As a setup relevant for quantum
computing, the special case of three qubits is considered for a Toffoli gate in
Sec.~\ref{sec3.1} and for a repeater relay station in Sec.~\ref{sec3.2}. An
approach to the navigation to a general target that does not rely on the
requirements of the previous sections is outlined in Sec.~\ref{sec4}. Finally, a
summary and outlook on future work is provided in Sec.~\ref{sec5}.
%-----------------------------------------------------------------------------
\section{
\label{sec2}
Bath-induced unitaries}
%-----------------------------------------------------------------------------
%-----------------------------------------------------------------------------
\subsection{
\label{sec2.1}
Zermelo navigation for bath-coupled systems}
\begin{figure}[ht]
\includegraphics[width=8.6cm]{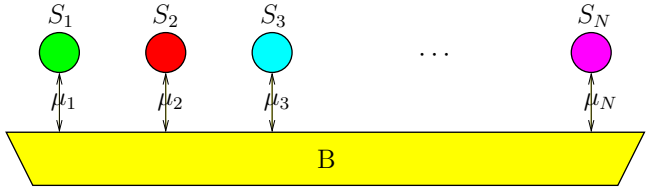}
\caption{\label{fig1}
A set of systems $S_j$ interacting with a shared bath (environment) B.
An extension with modulated system-bath interaction strengths, where 
each term in (\ref{HIj}) is multiplied with a $\mu_j(t)$ will be discussed in
Sec.~\ref{sec3} below.
}
\end{figure}

%-----------------------------------------------------------------------------
Consider $N$ systems $j$ $\!=$ $\!1,\ldots,N$ and one bath B as depicted in
Fig.~\ref{fig1}
%##############################################################################
%##############################################################################
with factorized individual system-bath interactions,
\begin{eqnarray}
\label{H}
  \hat{H}&=&\hat{H}_{\mathrm{S}}+\hat{H}_{\mathrm{B}}
  +\hat{H}_{\mathrm{SB}},
  \\
  \hat{H}_{\mathrm{S}}&=&\sum_{j=1}^N\hat{H}_j,
  \quad
  \hat{H}_{\mathrm{SB}}=\sum_{j=1}^N\hat{S}_j\otimes\hat{B}_j,
\label{HIj}
\end{eqnarray}
where $\hat{S}_j$ are Hermitian operators of system $j$
(hence $[\hat{S}_j,\hat{S}_k]$ $\!=$ $\!0$)
and $\hat{B}_j$ are Hermitian operators of the bath
(hence $[\hat{B}_j,\hat{B}_k]$ $\!\neq$ $\!0$).
%We will consider two cases:

%-----------------------------------------------------------------------------
%\subsection{Zermelo navigation}
%-----------------------------------------------------------------------------
The division in the quantum Zermelo navigation problem of the Hamiltonian
(\ref{ZH}) into an uncontrollable ``wind'' part $\hat{H}_0$ and a controllable
part $\hat{H}_{\mathrm{C}}$  is not fixed and refers to the operators one can
and wants to alter. In our case (\ref{H}) of a composite quantum system, a
natural assumption is that one can modify operators of the system but not those
of the bath. We may thus identify in (\ref{H}) 
\begin{equation}
  \hat{H}_0=\hat{H}_{\mathrm{S}},
  \quad
  \hat{H}_{\mathrm{C}}=\hat{H}_{\mathrm{B}}+\hat{H}_{\mathrm{SB}},
\end{equation}
for which (\ref{Zermelo}) with
$\hat{H}_{\mathrm{C}}(0)$ $\!\equiv$ $\!\hat{H}_{\mathrm{C}}$ becomes
\begin{equation}
\label{v1}
  \hat{H}_{\mathrm{C}}(t)
%  =\hat{U}_0(t)\hat{H}_{\mathrm{C}}(0)\hat{U}_0^{\dagger}(t)
  =\hat{H}_{\mathrm{B}}+\sum_{j=1}^N\hat{S}_j(t)\otimes\hat{B}_j,
\end{equation}
where 
%$\hat{U}_0(t)$ $\!=$ $\!\mathrm{e}^{-\mathrm{i}\hat{H}_0t}$,
$\hat{S}_j(t)$ $\!\equiv$ $\!\hat{U}_j(t)\hat{S}_j\hat{U}_j^\dagger(t)$ with
$\hat{U}_j(t)$ $\!\equiv$ $\!\mathrm{e}^{-\mathrm{i}\hat{H}_jt}$.
%and $\hat{B}_j(t)$
%$\!=$ $\!\hat{U}_{\mathrm{B}}(t)\hat{B}_j\hat{U}_{\mathrm{B}}^\dagger(t)$ where
We see that only the individual system operators $\hat{S}_j$ have to be
navigated in their respective local winds $\hat{H}_j$, whereas all bath
operators $\hat{H}_{\mathrm{B}}$ and $\hat{B}_j$ remain unaltered.

Given two general non-commuting and explicitly time-dependent Hamiltonians
$\hat{H}_1(t)$ and $\hat{H}_2(t)$, defining $\hat{U}_1(t_2,t_1)$ $\!\equiv$
$\!\mathrm{T}\mathrm{e}^{-\mathrm{i}\int_{t_1}^{t_2}\!\mathrm{d}t
\hat{H}_1(t)}$, and using the convention
$\hat{U}_1(t_1,t_2)$ $\!=$ $\!\hat{U}_1^\dagger(t_2,t_1)$, we can
disentangle the time-ordered product according to \cite{bookSchulman}
\begin{eqnarray}
  \hat{U}(t_2,t_1)\!\!\!&\equiv&\!\!\!\mathrm{T}\mathrm{e}^{-\mathrm{i}
  \!\!\int_{t_1}^{t_2}\!\mathrm{d}t[\hat{H}_1(t)+\hat{H}_2(t)]}
  \\
  \!\!\!&=&\!\!\!\hat{U}_1(t_2,t_1)\mathrm{T}\mathrm{e}^{-\mathrm{i}
  \!\!\int_{t_1}^{t_2}\!\mathrm{d}t\hat{U}_1(t_1,t)\hat{H}_2(t)\hat{U}_1(t,t_1)}
\label{s2}
  \\
  \!\!\!&=&\!\!\!\hat{U}_1(t_2,t_3)\mathrm{T}\mathrm{e}^{-\mathrm{i}
  \!\!\int_{t_1}^{t_2}\!\mathrm{d}t\hat{U}_1(t_3,t)\hat{H}_2(t)\hat{U}_1(t,t_3)}
  \hat{U}_1(t_3,t_1),
  \nonumber\\
\end{eqnarray}
where the auxiliary time $t_3$ can be chosen as desired.
Applying (\ref{s2}) with $t_1$ $\!=$ $\!0$, $t_2$ $\!=$ $\!t$,
$\hat{H}_1$ $\!=$ $\!\hat{H}_{\mathrm{S}}+\hat{H}_{\mathrm{B}}$,
$\hat{H}_2(\tau)$ $\!=$ $\!\sum_{j=1}^N\hat{S}_j(\tau)\otimes\hat{B}_j$ gives
for the total evolution operator $\hat{U}(t)$ $\!\equiv$ $\!\hat{U}(t,0)$
\begin{eqnarray}
  \hat{U}(t)&=&\mathrm{T}\mathrm{e}^{-\mathrm{i}
  \int_{0}^{t}\!\mathrm{d}\tau\hat{H}(\tau)}
  \nonumber\\
  &=&\hat{U}_{\mathrm{S}}(t)\hat{U}_{\mathrm{B}}(t)
  \hat{U}_{\mathrm{SB}}(t),
\label{U}
  \\
  \hat{U}_{\mathrm{SB}}(t)&=&
  \mathrm{T}\mathrm{e}^{-\mathrm{i}\int_{0}^{t}\!\mathrm{d}\tau
  \hat{H}_{\mathrm{SB}}(\tau)}
  \equiv\mathrm{e}^{-\mathrm{i}t\hat{H}_{\mathrm{eff}}(t)},
\label{USB}
\end{eqnarray}
where $\hat{U}_{\mathrm{S}}(t)$ $\!\equiv$
$\!\mathrm{e}^{-\mathrm{i}\hat{H}_{\mathrm{S}}t}$,
$\hat{U}_{\mathrm{B}}(t)$ $\!\equiv$
$\!\mathrm{e}^{-\mathrm{i}\hat{H}_{\mathrm{B}}t}$ and
$\hat{H}_{\mathrm{SB}}(\tau)$ $\!\equiv$
$\!\sum_{j=1}^N\hat{S}_j\otimes\hat{B}_j(\tau)$ with $\hat{B}_j(\tau)$
$\!\equiv$
$\!\hat{U}_{\mathrm{B}}^\dagger(\tau)\hat{B}_j\hat{U}_{\mathrm{B}}(\tau)$.
[In physical terms, this dynamic time dependence of
$\hat{H}_{\mathrm{SB}}(\tau)$ via $\hat{B}_j(\tau)$ refers to the Dirac
interaction picture and must be distinguished from the explicit time dependence
of $\hat{H}_{\mathrm{C}}(t)$ via $\hat{S}_j(t)$ introduced in (\ref{v1}). The
latter refers to the Schr\"odinger picture and describes the actual navigation
carried out by the experimenter.]

Eq.~(\ref{USB}) defines some Hermitian operator $\hat{H}_{\mathrm{eff}}$, which
can be expressed by means of the Magnus series expansion \cite{Mag54}.
$\hat{H}_{\mathrm{eff}}$ itself is time-dependent, and we have left a factor $t$
for dimensional reasons. Since
\begin{equation}
  [\hat{H}_{\mathrm{SB}}(t_1),\hat{H}_{\mathrm{SB}}(t_2)]=\sum_{j,k=1}^N
  \hat{S}_j[\hat{B}_j(t_1),\hat{B}_k(t_2)]\hat{S}_k,
\end{equation}
in the special case when $[\hat{B}_j(t_1),\hat{B}_k(t_2)]$ are $c$-number
functions, only the first two terms of the Magnus series are nonzero, and we
obtain a closed expression
\begin{eqnarray}
  \hat{H}_{\mathrm{eff}}(t)
  &=&\frac{1}{t}\int_{0}^{t}\mathrm{d}\tau
  \hat{H}_{\mathrm{SB}}(\tau)
  \nonumber\\
  &&-\frac{\mathrm{i}}{2t}
  \int_{0}^{t}\mathrm{d}t_1\int_{0}^{t_1}\mathrm{d}t_2
  [\hat{H}_{\mathrm{SB}}(t_1),\hat{H}_{\mathrm{SB}}(t_2)]\quad\quad
  \\
  &=&\sum_{j=1}^N\hat{S}_j\otimes\hat{F}_j(t)
  +\sum_{j,k=1}^N\hat{S}_j\kappa_{jk}(t)\hat{S}_k,
\label{exso}
\end{eqnarray}
where
\begin{eqnarray}
\label{Dneu}
  \hat{F}_j(t)&=&\frac{1}{t}\int_0^t\mathrm{d}\tau\hat{B}_j(\tau),
  \\
\label{kappaneu}
  \kappa_{jk}(t)&=&-\frac{\mathrm{i}}{2t}
  \int_0^t\!\mathrm{d}t_1\int_0^{t_1}\!\mathrm{d}t_2
  [\hat{B}_j(t_1),\hat{B}_k(t_2)].
\end{eqnarray}
%-----------------------------------------------------------------------------
\subsection{
\label{sec2.2}
Linear coupling to a bosonic bath}
%-----------------------------------------------------------------------------
A simple example in which the Magnus series breaks off at second order is a
bath of bosonic modes \cite{Dur11a,Nir11,Dur11b}. The bath parts of
$\hat{H}_{\mathrm{SB}}$ in (\ref{HIj})
\begin{equation}
\label{A10}
  \hat{B}_j=\sum_l{A}_{jl}\hat{b}_l+h.a.
\end{equation}
are linear combinations of the bosonic mode operators $\hat{b}_l$
and $h.a.$ denotes the Hermitian conjugate. A bath Hamiltonian
\begin{equation}
\label{bh}
  \hat{H}_{\mathrm{B}}=\sum_l
  \omega_l\,\hat{b}_l^\dagger\hat{b}_l
\end{equation}
then gives the time evolution
$\hat{B}_j(t)$ $\!=$
$\!\sum_l{A}_{jl}\mathrm{e}^{-\mathrm{i}\omega_l{t}}\hat{b}_l$ $\!+$ $\!h.a.$
Using $[\hat{b}_l,\hat{b}_{l^\prime}^\dagger]$ $\!=$ $\!\delta_{ll^\prime}$ we
can verify that the above-mentioned $c$-number condition is fulfilled,
$[\hat{B}_j(t_1),\hat{B}_k(t_2)]$ $\!=$ $\!2\mathrm{i}\,\mathrm{Im}
\sum_l{A}_{jl}{A}_{kl}^*\mathrm{e}^{-\mathrm{i}\omega_l(t_1-t_2)}$. If the 
${A}_{jl}$ couple only to harmonics of some given frequency $\omega$, i.e.,
they are nonzero only for $\omega_l$ $\!=$ $\!l\omega$, $l$ $\!\in$
$\!\mathbb{N}$, then for discrete times
\begin{equation}
\label{tm}
  t_m=m\Delta{t},\quad\Delta{t}=\frac{2\pi}{\omega}
\end{equation}
($m$ $\!\in$ $\!\mathbb{N}$) the integral (\ref{Dneu}) vanishes,
$\hat{F}_j(t_m)$ $\!=$ $\!0$, whereas (\ref{kappaneu}) reduces to
$\kappa_{jk}$ $\!\equiv$ $\!\kappa_{jk}(t_m)$ $\!=$
$\!-\mathrm{Re}\sum_l{A}_{jl}{A}_{kl}^*/(l\omega)$. Combining the ${A}_{jl}$ to
a matrix $\bm{A}$ and defining a diagonal matrix $\bm{\Omega}$ with elements
${\Omega}_{ll^\prime}$ $\!=$ $\!-\delta_{ll^\prime}/(l\omega)$ we can write
$\kappa_{jk}$ $\!=$ $\!\mathrm{Re}(\bm{A}\bm{\Omega}\bm{A}^\dagger)_{jk}$
and combine the $\kappa_{jk}$ to a matrix $\bm{\kappa}$.

In summary, the total evolution operator (\ref{U}) factorizes at the discrete
times $t_m$ according to
\begin{equation}
\label{final}
  \hat{U}(m\Delta{t})
  =\mathrm{e}^{-\mathrm{i}m\Delta{t}\hat{H}_{\mathrm{S}}}
  \mathrm{e}^{-\mathrm{i}m\Delta{t}
  \bm{\hat{S}}\cdot\bm{\kappa}\cdot\bm{\hat{S}}}
  \otimes\mathrm{e}^{-\mathrm{i}m\Delta{t}\hat{H}_{\mathrm{B}}},
\end{equation}
where $\bm{\hat{S}}\cdot\bm{\kappa}\cdot\bm{\hat{S}}$ $\!=$
$\!\sum_{j,k=1}^N\hat{S}_j\kappa_{jk}\hat{S}_k$. The first two exponentials are
a result of the Zermelo navigation, whereas the last one describes the free
evolution of the decoupled bath.
%-----------------------------------------------------------------------------
\subsection{
\label{sec2.3}
Speed limit}
%-----------------------------------------------------------------------------
In our scenario, a set of $N$ systems interact with a common bath such that the
overall unitary evolution factorizes for discrete times (\ref{tm}) into system
and bath parts. The bath nevertheless affects the system evolution by means of
the exponential containing the $\hat{S}_j$ in (\ref{final}). This is in contrast
to Markovian bath effects \cite{bookBreuer}, where information dissipates
irreversibly into the environment thus preventing the realization of a unitary
transformation. Unlike the system evolution, the evolution of the bath itself in
(\ref{final}) is not affected.
The Zermelo navigation serves to remove any time dependence from the
control Hamiltonian in the interaction picture which allows a straightforward
evaluation of the obtained target unitary by means of the Magnus identity.
Due to the need of a well-defined coupling, the bath here acts as an ancillary
system such as a field confined by a cavity rather than an uncontrolled
environment.

Obviously, $\mathrm{Tr}[\hat{H}_{\mathrm{C}}^2(t)]$ $\!=$
$\!\mathrm{Tr}_1\ldots\mathrm{Tr}_N\mathrm{Tr}_{\mathrm{B}}
[\hat{H}_{\mathrm{C}}^2(t)]$ is undetermined but fixed because it is
time-independent. In contrast to the
finite-dimensional systems for which the quantum Zermelo navigation problem has
been discussed so far, the bath considered here is an infinite set of
infinite-dimensional systems. The Zermelo solution of the problem does not
contain any reference to the system size, however, which suggests that it holds
in our more general setting too, cf. also a comment in \cite{Rus14b}.
(We may truncate the bath with respect to the number of relevant bath modes and
at some sufficiently high occupation number for each of these modes without
changing any physical properties of the model.)
In (\ref{final}), the infinite bath is decoupled from the unitary evolution of
the $N$ systems, where the latter may be finite or infinite dimensional.
Restricting the modulation (imposed time-dependency) to the systems renders the
method independent on any need to manipulate the bath.

Since we apply Zermelo navigation, the described method of bath-induced
realization of the above type of unitaries is time-optimal with respect to the
local ``winds'' $\hat{H}_j$. Substituting {\small $\omega$ $\!=$
$\!\frac{2E_0}{\hbar}$}, where $E_0$ is the ground state energy of a harmonic
oscillator [i.e., if we set {\small $\hat{H}_{\mathrm{B}}$ $\!\equiv$
$\!\hbar\omega\bigl(\hat{n}$ $\!+$ $\!\frac{1}{2}\bigr)$}], the minimum time
required for refactorization of the unitary into system and bath parts can be
written in the canonical form mentioned in the beginning
\begin{equation}
  t_{\mathrm{min}}=\frac{\pi\hbar}{E_0}.
\end{equation}
A single harmonic oscillator with fundamental frequency $\omega$ here serves as
a minimum system suitable to act as bath in the described way.
This minimal implementation time is finite but small for large $E_0$. An
observer on a large time scale, on which she cannot resolve $t_{\mathrm{min}}$,
may thus be unaware of it and even the existence of a background bath. Such an
observer would simply witness a time-continuous evolution of the systems, whose
bath-mediated interactions would exhibit as direct pairwise system interactions.
%-----------------------------------------------------------------------------
\section{
\label{sec3}
Concatenations}
%-----------------------------------------------------------------------------
So far we have ignored the problem of reaching a given target unitary.
In \cite{Bro14a}, a method is presented to calculate the initial
$\hat{H}_{\mathrm{C}}(0)$ such that $\hat{U}(t)$ after some time $t$ reaches the
target $\hat{U}_{\mathrm{F}}$, $\hat{U}(t)$ $\!=$
$\!\hat{U}_{\mathrm{F}}\hat{U}_{\mathrm{I}}^{\dagger}$, so that
$\hat{U}(0)\hat{U}_{\mathrm{I}}$ $\!=$ $\!\hat{U}_{\mathrm{I}}$ and 
$\hat{U}(t)\hat{U}_{\mathrm{I}}$ $\!=$ $\!\hat{U}_{\mathrm{F}}$.
In our context, we face the additional restriction to pairwise couplings
$\bm{\hat{S}}\cdot\bm{\kappa}\cdot\bm{\hat{S}}$ in (\ref{final}).
The system part of (\ref{final}) is itself a product of two factors,
$\hat{U}_{\mathrm{system}}(m\tau)$ $\!=$
$\!\mathrm{e}^{-\mathrm{i}m\tau\hat{B}}\mathrm{e}^{-\mathrm{i}m\tau
\hat{A}}$, given by $\hat{A}$ $\!=$
$\!\sum_{j,j^\prime=1}^N\hat{S}_j\kappa_{jj^\prime}\hat{S}_{j^\prime}$ and
$\hat{B}$ $\!=$ $\!\sum_{j=1}^N\hat{H}_j$, respectively, where $\tau$ is fixed
by (\ref{tm}). 
This restricts the set of implementable $N$-system target unitaries
$\hat{U}_{\mathrm{system}}(t_m=m\tau)$ $\!\stackrel{!}{=}$
$\!\hat{U}_{\mathrm{F}}$, even if the $\hat{S}_j$ can be chosen at will.
To generalize the set of reachable $\hat{U}_{\mathrm{F}}$, we may concatenate
individual Zermelo navigations by adjusting at times
$t_m$ $\!=$ $\!\sum_{k=1}^m\tau_k$ [by means of $\omega$, cf. (\ref{tm})]
$\tau$ to $\tau_{m+1}$, $\kappa_{jj^\prime}$ to $\kappa_{jj^\prime}^{(m+1)}$
[by multiplying each term in $\hat{H}_{\mathrm{SB}}$ in (\ref{HIj}) with a
system-bath interaction strength $\mu_j^{(m+1)}$ $\!\ge$ $\!0$], and reset the
$\hat{S}_j(t)$ to $\hat{S}_j$. This gives $\hat{U}_{\mathrm{system}}(t_m)$ $\!=$
$\!\prod_{k=1}^m\mathrm{e}^{-\mathrm{i}\beta_k\hat{B}}
\mathrm{e}^{-\mathrm{i}\alpha_k\hat{A}_k^{(\mathcal{N})}}$ with 
$\hat{A}_k^{(\mathcal{N})}$ $\!=$ $\!\|\hat{A}_k\|^{-1}\hat{A}_k$,
$\alpha_k$ $\!=$ $\!\|\hat{A}_k\|\tau_k$, $\beta_k$ $\!=$ $\!\tau_k$, and given
$\hat{A}_k$ $\!=$ $\!\sum_{j,j^\prime=1}^N\hat{S}_j\kappa_{jj^\prime}^{(k)}
\hat{S}_{j^\prime}$ with $\kappa_{jj^\prime}^{(k)}$ $\!=$
$\!\mu_j^{(k)}\mu_{j^\prime}^{(k)}(\tau_k/\tau)\kappa_{jj^\prime}$. Here, we
factored out $\|\hat{A}\|^2$ $\!\equiv$ $\!\mathrm{Tr}(\hat{A}^\dagger\hat{A})$
for convenience, and without loss of generality we assume that
$\|\hat{B}\|$ $\!=$ $\!1$.
The special case $\mu_j^{(k)}$ $\!\equiv$ $\!\mu^{(k)}$ and hence
$\kappa_{jj^\prime}^{(k)}$ $\!\sim$ $\!\kappa_{jj^\prime}$
generates an alternate product for which $\hat{A}_k^{(\mathcal{N})}$ $\!\equiv$
$\!\hat{A}^{(\mathcal{N})}$, which can approximate any
$\hat{U}_{\mathrm{F}}$ $\!=$ $\!\mathrm{e}^{-\mathrm{i}\hat{H}}$, with $\hat{H}$
being a member of the algebra spanned by the multicommutators from
$\hat{A}$ and $\hat{B}$ \cite{lloyd2}, where the $\alpha_k$ and $\beta_k$ can be
determined from $\hat{U}_{\mathrm{F}}$ \cite{akulin2}.
The resulting concatenated navigation represents a (piecewise differentiable)
evolution from $\hat{I}$ to $\hat{U}_{\mathrm{F}}$ under the constraint of fixed
$\hat{A}^{(\mathcal{N})}$ (or $\hat{A}_k^{(\mathcal{N})}$) and $\hat{B}$.
The Zermelo-optimality holds only for the segments $\mathrm{e}^{-\mathrm{i}
\beta_k\hat{B}}\mathrm{e}^{-\mathrm{i}\alpha_k\hat{A}_k^{(\mathcal{N})}}$,
during which the two factors
$\mathrm{e}^{-\mathrm{i}\alpha_k\hat{A}_k^{(\mathcal{N})}}$ and
$\mathrm{e}^{-\mathrm{i}\beta_k\hat{B}}$ are hence implemented at once within
$\tau_k$ rather than in consecution, sparing the need to switch between the
generators $\hat{A}_k$ and $\hat{B}$.

The number $N$ of systems can be arbitrary including infinite.
In the fundamental case $N$ $\!=$ $\!1$, in which no division into subsystems is
assumed, the effect of the bath reduces to ``dressing'' the
system evolution, $\hat{A}_k$ $\!\sim$ $\!\hat{S}^2$. For $N$ $\!>$ $\!1$, 
the individual systems $j$ and their bath-coupling as defined by $\hat{H}_j$
and $\hat{S}_j$ can be set arbitrarily and do not need to be the same for each
$j$. For $N$ $\!>$ $\!2$, one may argue that even if we can choose the
$\hat{S}_j$ at will, they are local operators as are the $\hat{H}_j$. However,
since $\hat{A}$ (or $\hat{A}_k$) is 2-local (i.e., it represents a pairwise
system-coupling), and each commutator of a $k$-local operator with $\hat{A}$
generally yields a $(k$ $\!+$ $\!1)$-local operator unless the number of systems
has been reached, the concatenation builds up $N$-system unitaries, analogous to
a quantum circuit consisting of one- and two-qubit gates.
%-----------------------------------------------------------------------------
\subsection{
\label{sec3.1}
Quantum circuits with qubits}
%-----------------------------------------------------------------------------
A ``continuous variable'' example is a set of harmonic oscillators with
parametric-amplifier type couplings, $\hat{S}_j=\hat{n}_j$, for which the bath
induces Kerr and cross-Kerr nonlinearities
$\hat{S}_j\hat{S}_k=\hat{n}_j\otimes\hat{n}_k$. Of special interest for quantum
information processing are two-level systems, however, for which we now provide
some explicit examples. If $\hat{H}_0$ and $\hat{H}_{\mathrm{C}}$ are given
appropriately, $\hat{U}_{\mathrm{F}}$ may represent a desired gate without need
of concatenations. An example for $N$ $\!=$ $\!2$, where $\hat{H}_0$ and
$\hat{H}_{\mathrm{C}}$ commute, is the CNOT-gate
\cite{bookBraunstein} 
$\hat{U}_{\mathrm{CNOT}}$ $\!=$ $\!
\mathrm{e}^{-\mathrm{i}\frac{\pi}{4}}
\mathrm{e}^{\mathrm{i}\frac{\pi}{4}\hat{\sigma}_1^{(2)}}
\mathrm{e}^{\mathrm{i}\frac{\pi}{4}\hat{\sigma}_3^{(1)}}
\mathrm{e}^{-\mathrm{i}\frac{\pi}{4}\hat{\sigma}_3^{(1)}\hat{\sigma}_1^{(2)}}$,
which can be implemented if
$\hat{S}_1$ $\!=$ $\!-\hat{H}_1$ $\!=$ $\!\hat{\sigma}_3^{(1)}$,
$\hat{S}_2$ $\!=$ $\!-\hat{H}_2$ $\!=$ $\!\hat{\sigma}_1^{(2)}$,
$\tau$ $\!=$ $\!\frac{\pi}{4}$, and $\kappa_{12}$ $\!=$ $\!\frac{1}{2}$.

If the $\hat{H}_j$ and $\hat{S}_j$ are fixed by the experimental setup,
concatenations are required to implement a desired $\hat{U}_{\mathrm{F}}$.
As a concrete example, we consider three qubits with randomly given
$\hat{H}_j$ and $\hat{S}_j$, whose interaction strengths with a shared bosonic
bath can be altered [by multiplying each term in $\hat{H}_{\mathrm{SB}}$ in
(\ref{HIj}) with a factor $\mu_j^{(k)}$ $\!\ge$ $\!0$ over time interval
$\tau_k$ as mentioned above], and where the bosonic bath is eliminated by
Zermelo-navigation of the $\hat{S}_j$ leading to products
$\hat{U}_{\mathrm{system}}(t_m)$ $\!=$
$\!\prod_{k=1}^m\mathrm{e}^{-\mathrm{i}\tau_k\hat{B}}\mathrm{e}^{-\mathrm{i}
\tau_k\|\hat{A}_k\|\hat{A}_k^{(\mathcal{N})}}$ for the $N$-system unitary
as described.
The protocols are thus sequences of these finite time periods $\tau_k$, over
which the system-bath interaction strengths of the qubits $j$ are multiplied by
the respective $\mu_j^{(k)}$. Specifically, we consider (a) a synchronous
protocol, $\mu_j^{(k)}$ $\!\equiv$ $\!\mu^{(k)}$, generating an alternate
product as mentioned above. An alternative is (b) an asynchronous protocol
consisting of concatenations of 8-step cycles, where all subsets of the qubits
are brought in contact with the bath as follows: over $\tau_1$, all qubits are
detached from the bath, $\mu_j^{(1)}$ $\!=$ $\!0$, after which qubit 1, then
qubit 2, and then qubit 3 alone is brought in contact with the bath over
$\tau_2$, $\tau_3$, $\tau_4$, respectively. After this, only qubits 1 and 2,
then only qubits 1 and 3, and then only qubits 2 and 3 are brought
simultaneously in contact with the bath over $\tau_5$, $\tau_6$, $\tau_7$,
respectively. Finally, all 3 qubits are brought in contact with the bath over
$\tau_8$. After this, at time $t_8$ $\!=$ $\!\sum_{k=1}^8\tau_k$, the cycle is
repeated from the beginning over different times $\tau_9\cdots\tau_{16}$, and
with different interaction strengths for those qubits in bath contact, and so
forth, until a maximum number $n$ of exponential factors in
$\hat{U}_{\mathrm{system}}$ has been reached. The qubit-bath interaction
strengths of those qubits $j$ which are in bath contact over $\tau_k$ are
adjusted synchronously, $\mu_j^{(k)}$ $\!\equiv$ $\!\mu^{(k)}$ $\!>$ $\!0$ as
in (a). In both protocols (a) and (b), we minimize the (squared) operator
distance $D$ $\!=$ $\!\|\hat{U}_{\mathrm{system}}-\hat{U}_{\mathrm{F}}\|^2$
\cite{Rab05} of the implemented unitary $\hat{U}_{\mathrm{system}}$ and a given
target $\hat{U}_{\mathrm{F}}$ by gradient descent. The minimum distance reached
for a given number of exponential factors $n$ is shown in Fig.~\ref{fig2} for
the case where $\hat{U}_{\mathrm{F}}$ is either the CNOT (see Sec.~\ref{sec3.2}
below) or the Toffoli (i.e., C$^2$NOT) gate \cite{bookNielsen} for both types of
protocols.
%##############################################################################
\begin{figure}[ht]
\includegraphics[width=8.6cm]{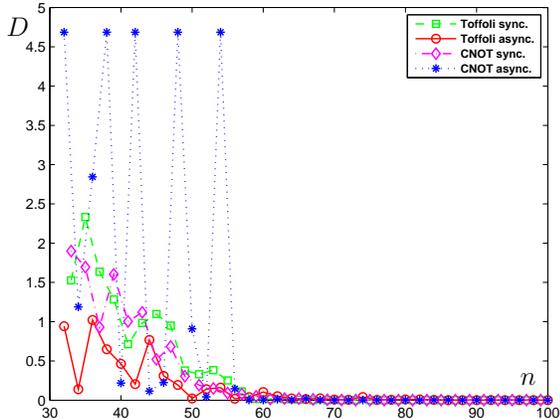}
\caption{\label{fig2}
 %(color online).
Minimized squared operator distance
$D$ $\!=$ $\!\|\hat{U}_{\mathrm{system}}-\hat{U}_{\mathrm{F}}\|^2$ as a function
of the number $n$ of exponential factors in $\hat{U}_{\mathrm{system}}$ for
concatenated Zermelo navigations with three qubits $j$ $\!=$ $\!1,2,3$ sharing
a bosonic bath. The plots show the implementation of a $\hat{U}_{\mathrm{F}}$ =
CNOT gate between qubits 1 and 3 as well as a $\hat{U}_{\mathrm{F}}$ = Toffoli
gate, using the synchronous and the asynchronous protocol. The CNOT gate refers
to the quantum repeater relay station depicted in Fig.~\ref{fig5}. The local
winds $\hat{H}_j$ and couplings $\hat{S}_j$ are randomly given for each
individual $j$.
}
\end{figure}
%##############################################################################
The choice of a Toffoli gate is motivated by its relevance for quantum
computing, but analogous results can be obtained for a random gate.
Fig.~\ref{fig2} demonstrates that a desired target $\hat{U}_{\mathrm{F}}$ can
generally be reached for both types of protocols as soon as $n$ surpasses a
threshold [$>(2^3)^2$ $\!=$ $\!64$, albeit smaller $n$ may suffice for
specifically given $\hat{H}_j$, $\hat{S}_j$, and $\hat{U}_{\mathrm{F}}$ as in
the closed expression for $\hat{U}_{\mathrm{CNOT}}$ mentioned above]. An example
for each type of protocol is illustrated in Fig.~\ref{fig3}. The choice of a
synchronous protocol for the Toffoli gate and an asynchronous protocol for the
CNOT gate is here irrelevant, since each gate can be generated with both types
of protocols.
%##############################################################################
\begin{figure}[ht]
\includegraphics[width=8.6cm]{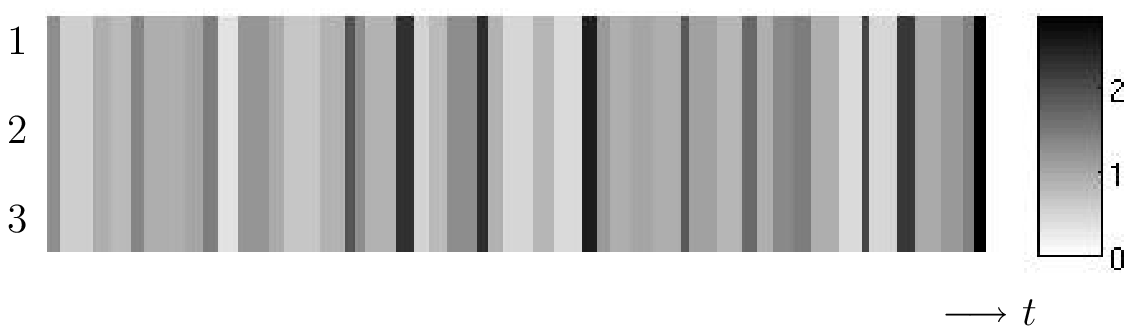}
\includegraphics[width=8.6cm]{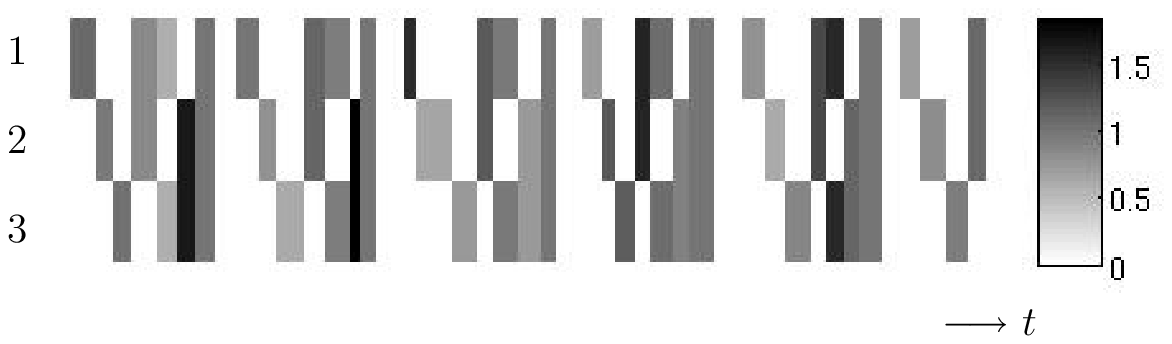}
\caption{\label{fig3}
Illustration of example protocols consisting of concatenated Zermelo navigations
of qubits 1,2,3. Shown are the strengths $\|\hat{A}_k\|$ of the system-bath
interaction (the navigator's ``engine power'') over the time intervals $\tau_k$
for those qubits in bath contact during $\tau_k$.
Upper plot: synchronous protocol generating a Toffoli gate ($n$ $\!=$ $\!92$),
lower plot: asynchronous protocol generating a CNOT gate between qubit 1 and 3
($n$ $\!=$ $\!84$). The CNOT gate refers to the quantum repeater relay station
shown in Fig.~\ref{fig5}.
}
\end{figure}
%##############################################################################

\begin{figure}[ht]
\includegraphics[width=8.6cm]{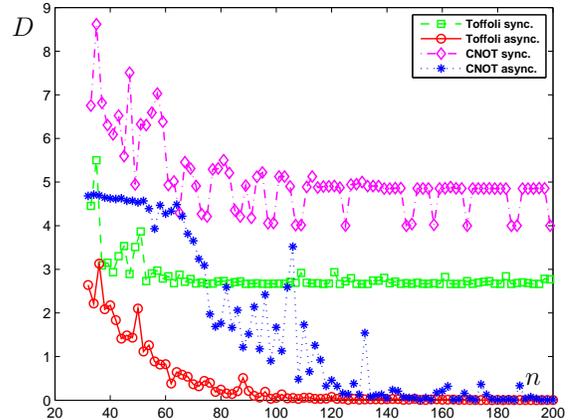}
\caption{\label{fig4}
 %(color online).
Same analysis as in Fig.~\ref{fig2}, except that the randomly chosen local winds
$\hat{H}_j$ and couplings $\hat{S}_j$ are identical for each $j$.
}
\end{figure}

The asynchronous protocol requires a system-resolved switching of the
coupling but exhibits a more robust convergence. This is illustrated in
Fig.~\ref{fig4}, where in contrast to Fig.~\ref{fig2}, the three qubits and
their couplings are identical, i.e., there is one single (randomly given)
$\hat{H}$, so that the local winds (acting in the three-qubit Hilbert space) are
$\hat{H}_1$ $\!=$ $\!\hat{H}\otimes\hat{I}\otimes\hat{I}$,
$\hat{H}_2$ $\!=$ $\!\hat{I}\otimes\hat{H}\otimes\hat{I}$,
$\hat{H}_3$ $\!=$ $\!\hat{I}\otimes\hat{I}\otimes\hat{H}$, and the local
couplings $\hat{S}_j$ are analogously given by one single (randomly given)
$\hat{S}$. Fig.~\ref{fig4} demonstrates that the synchronous protocol fails in
this case, in contrast to the asynchronous protocol.
%##############################################################################
%##############################################################################
%-----------------------------------------------------------------------------
\subsection{
\label{sec3.2}
Quantum repeater relay station}
\begin{figure}[ht]
\includegraphics[width=8.6cm]{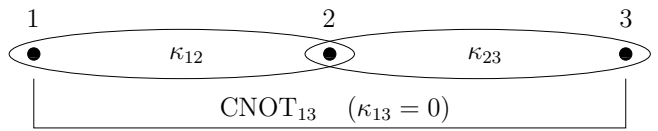}
\caption{\label{fig5}
Quantum repeater relay station: qubits 1 and 3 are coupled to qubit 2 but not
to each other directly. The intermediate qubit 2 can be eliminated by
concatenated Zermelo navigations to generate a direct coupling between qubits 1
and 3. An example protocol for generating a CNOT gate between these qubits is
shown in the lower plot of Fig.~\ref{fig3}. Analogous results can be obtained
for a random gate.
}
\end{figure}

%-----------------------------------------------------------------------------
Depending on the application, the system coupling may follow a restricted
topology. For example, the systems may be arranged in a chain, and bath-induced
coupling is possible only for nearest neighbors. In the simplest case, three
qubits form a linear array as shown in Fig.~\ref{fig5}, with a presence of
direct couplings $\kappa_{12}$ and $\kappa_{23}$, but absence of an end-to-end
coupling, $\kappa_{13}$ $\!=$ $\!0$.
%##############################################################################
%##############################################################################
This setup resembles our original problem with the role of the bath adopted by
the intermediate qubit 2. Similar to the elimination of the bath by Zermelo
navigation, the intermediate qubit 2 can be eliminated by concatenated Zermelo
navigations to generate a (qubit 2 - mediated) direct coupling between qubits 1
and 3 alone and thus a desired target gate $\hat{U}_{\mathrm{F}}^{(13)}$. Since
nesting of this procedure on larger time scales for a successive elimination of
blocks of subsystems is (aside from the open system dynamics) reminiscent of
quantum repeaters \cite{Bri98,San11,Dia15}, we refer to the setup depicted in
Fig.~\ref{fig5} as a quantum repeater relay station.
%-----------------------------------------------------------------------------
\section{
\label{sec4}
Generalizations}
%-----------------------------------------------------------------------------
%-----------------------------------------------------------------------------
\subsection{
\label{sec4.1}
Motivation}
%-----------------------------------------------------------------------------
Although concatenating Zermelo navigations as discussed in Sec.~\ref{sec3} is a
straightforward option for reaching a target, similar to the construction of a
quantum circuit from individual gates, the concatenations as a whole are no
longer time-optimal, since they introduce refactorizations of the overall
system-bath unitary at the intermediate times
$t_m$ $\!=$ $\!\sum_{k=1}^m\tau_k$ as an artifact of this approach. The optimal
system-bath trajectory would be a brachistochrone $\hat{U}(t)$ that starts at
the identity and refactorizes only at the time of arrival.
Even if we choose (concatenated) Zermelo navigations, we must ensure (a) the
experimental realization of the required modulations $\hat{S}_j(t)$ (and their
resets), (b) the presence of a linear coupling to a bosonic bath with the
desired exclusive coupling to harmonics of a fundamental frequency $\omega$
(and its adjustment), and (c) detailed knowledge of all operators $\hat{H}_j$,
$\hat{S}_j$, and coupling coefficients $\kappa_{jj^\prime}$, along with an
absence of operator noise and decoherence. The realization of these requirements
can be challenging.

To avoid these problems, we now consider an alternative that relies on
measurements alone, i.e.,  a ``closed loop'' - scheme.  Rather than minimizing
the squared operator distance $D$, we maximize the fidelity \cite{Ped07}, which
is in our context defined as an average overlap $F$ $\!\equiv$
$\!\overline{\mathrm{Tr}\bigl(\hat{\varrho}_{\mathrm{out}}\hat{\Pi}\bigr)}$.
Here, a factorized system-bath input state $\hat{\varrho}_{\mathrm{in}}$ $\!=$
$\!|\Psi\rangle\langle\Psi|\otimes\hat{\varrho}_{\mathrm{B}}$ is transformed to
$\hat{\varrho}_{\mathrm{out}}$ $\!=$
$\!\hat{U}\hat{\varrho}_{\mathrm{in}}\hat{U}^\dagger$ by a total system-bath
unitary operation $\hat{U}$ describing the device, after which an overlap is
measured with a projector $\hat{\Pi}$ $\!=$
$\!\hat{U}_{\mathrm{F}}|\Psi\rangle\langle\Psi|\hat{U}_{\mathrm{F}}^\dagger$.
This projector is determined by the respective input system state $|\Psi\rangle$
transformed by the desired target system unitary $\hat{U}_{\mathrm{F}}$ and
defines the measurement device. $\overline{(\cdots)}$ denotes the
uniform average over all system states $|\Psi\rangle$. Since the fidelity can
thus be estimated from repeated binary measurements with sampled input states
$|\Psi\rangle$ and gradually improved, this approach is independent of any
assumptions about or knowledge of the environment and its coupling to the 
system.

\begin{figure}[ht]
\includegraphics[width=8.6cm]{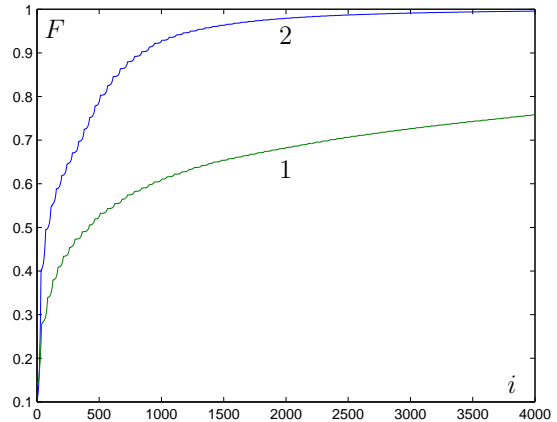}
\caption{\label{fig6}
 %(color online).
Simulation of a closed loop gradient ascent of the fidelity $F$ $\!=$
$\!\overline{\mathrm{Tr}\bigl[\hat{U}(|\Psi\rangle\langle\Psi|\otimes
\hat{\varrho}_{\mathrm{B}})\hat{U}^\dagger\hat{\Pi}\bigr]}^{(\mathcal{P})}$,
which can be estimated experimentally from sampled binary measurements of
$\hat{\Pi}$ $\!=$
$\!\hat{U}_{\mathrm{F}}|\Psi\rangle\langle\Psi|\hat{U}_{\mathrm{F}}^\dagger$,
with the number of iterations $i$ for three qubits sharing a random bath.
Plot (1): $\hat{U}_{\mathrm{F}}$ = Toffoli gate acting on the whole three-qubit
space, $\mathcal{P}$ $\!=$ $\!\mathcal{I}$, 
Plot (2): $\hat{U}_{\mathrm{F}}$ = Hadamard gate acting on a dynamically encoded
decoherence free subspace $\mathcal{P}$ = span\{$|000\rangle$, $|111\rangle$\},
using an asynchronous protocol with $n$ $\!=$ $\!576$ time steps as controls.
}
\end{figure}

As an illustration, we consider again the task of implementing a Toffoli gate as
shown in Fig.~\ref{fig6}. In contrast to Fig.~\ref{fig2}, the three qubits
$j$ $\!=$ $\!1,2,3$ (again with randomly given $\hat{H}_j$ and $\hat{S}_j$) are
now coupled to a common random environment (with randomly given
$\hat{H}_{\mathrm{B}}$ and $\hat{B}_j$) rather than the described bosonic bath.
Instead of Zermelo navigation, subsets of the qubits are brought in contact with
the environment for fixed time periods $\tau_k$ $\!\equiv$ $\!\tau$
in a fixed order given by the asynchronous protocol described above.
The controls are the system-environment interaction strengths
(cf. the $\mu_j^{(k)}$ above) over these fixed time periods. (Equivalently, we
may fix the system-environment interaction strength and tune the time periods
$\tau_k$ instead.) Plot 1 in Fig.~\ref{fig6} shows a simulation of the gradual
increase of the fidelity as it would be seen in a control loop.
%##############################################################################
%##############################################################################
For simplicity, we have assumed that during each iteration $i$, a reliable
estimate of the fidelity has been obtained by a sufficient number of
measurements. Depending on the respective environment and the number of
controls (time steps), such a control loop cannot guarantee to implement a
desired gate in practice, as is illustrated in plot 1 of Fig.~\ref{fig6} for the
Toffoli gate. Here, we have restricted ourselves to a 3-level ancillary system
initially in a maximally mixed state
$\hat{\varrho}_{\mathrm{B}}$ $\!=$ $\!\frac{1}{3}\hat{I}$ serving as
environment, whose properties are unknown to the control loop. Since with an
idealized Markovian bath (i.e., a bath in a narrower sense) no change of the
fidelity would be observable by definition, such a control loop can at the same
time serve to measure - in terms of the change of $F$ - the deviation of the
environment from an idealized bath and thus to witness the presence of
non-Markovian properties. The latter application does not require the
realization of a unitary with perfect (unit) fidelity.
%-----------------------------------------------------------------------------
\subsection{
\label{sec4.2}
Encoding a minimal noise subspace}
%-----------------------------------------------------------------------------
To mitigate or remedy this controllability problem, we may restrict ourselves to
input states $|\Psi\rangle\in\mathcal{P}$ belonging to a given subspace
$\mathcal{P}$. The fidelity $F$ $\!=$ $\!\overline{\mathrm{Tr}
\bigl(\hat{\varrho}_{\mathrm{out}}\hat{\Pi}\bigr)}^{(\mathcal{P})}$ is now given
by the uniform average $\overline{(\cdots)}^{(\mathcal{P})}$ with all states
$|\Psi\rangle\in\mathcal{P}$ only. $\mathcal{P}$ thus adopts the role of a
minimal noise or decoherence free subspace \cite{Lid14,Wan13}, which is
dynamically encoded by the control loop, without knowledge of any symmetries in
the open system dynamics. Plot 2 in Fig.~\ref{fig6} illustrates this in the
example of a Hadamard gate encoded in
$\mathcal{P}$=span\{$|000\rangle$, $|111\rangle$\}. Due to the
absence of any symmetries, the improvement of achievable fidelity is here due to
the redundancy introduced by reducing the dimension of the (logical) quantum
channel from 8 to 2.

The asynchronous protocol is here used merely in connection to Sec.~\ref{sec3},
where the piecewise constant $\mu_j$ are a consequence of the concatenated
Zermelo navigations. It is obvious that instead of this, three independent
continuous time-dependencies $\mu_j(t)$ can be applied just as well for the
maximization of the fidelity, cf. e.g. \cite{clausen18}.
%-----------------------------------------------------------------------------
\section{
\label{sec5}
Summary and outlook}
%-----------------------------------------------------------------------------
In summary, we have applied the solution for the quantum Zermelo navigation
problem to a scenario where a set of open systems share a bath. We have shown
that Zermelo navigation allows to generalize an analytic case of system-bath
factorization of the total unitary evolution operator to the presence of
arbitrary background fields $\hat{H}_0$, where only the individual system parts
$\hat{S}_j$ in the coupling operators have to be navigated in their respective
local winds $\hat{H}_j$, whereas all bath operators remain unaltered. We have
given a quantum state independent refactorization time limit in terms of the
minimal bath energy. Finally we addressed the reachability of general target
unitaries by concatenations of Zermelo navigations, making use of the
two-exponential form of the individual Zermelo unitaries, and demonstrated the
feasibility in numerical examples. In addition, to address the navigation to a
general target in an unknown environment, we have considered a measurement-based
closed loop scheme.

\begin{figure}[ht]
\includegraphics[height=3.2cm]{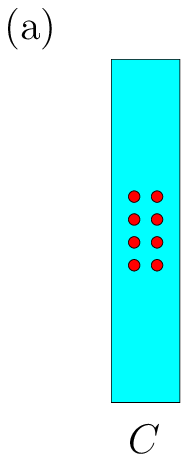}\hspace{1cm}
\includegraphics[width=6cm]{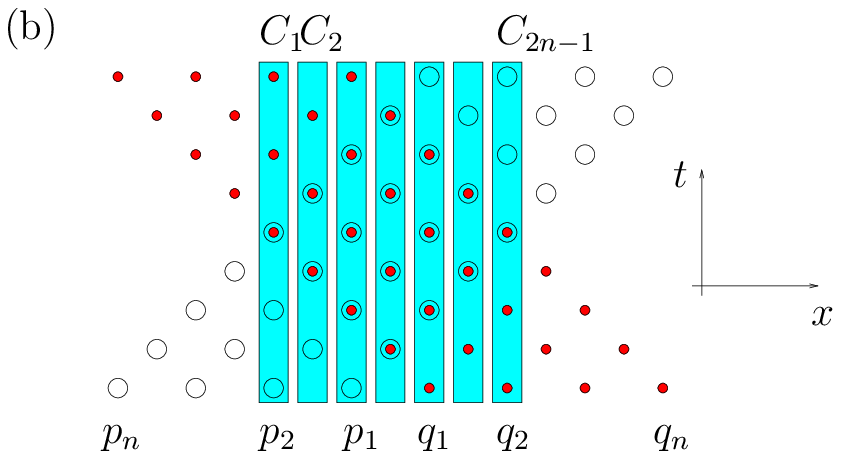}
\caption{\label{fig7}
 %(color online).
Possible schemes for controlled bath coupling: (a) static setup in which the
systems (circles) are arranged in an array confined by a cavity $C$;
(b) dynamic setup with two counterpropagating sequences of $n$ $\!=$ $\!N/2$
systems $p$ (circles) and $q$ (dots) which interact during coincident passages
through an array of $2n$ $\!-$ $\!1$ cavities: for identical interactions
$\kappa_{jk}$ within the cavities, the term
$\sum_{j,k=1}^N\hat{S}_j\kappa_{jk}\hat{S}_k$ then reduces to a
collective coupling $\sum_{j,k=1}^n\hat{p}_j\hat{q}_k$ $\!=$ $\!\hat{P}\hat{Q}$,
where $\hat{P}$ $\!=$ $\!\sum_{j=1}^n\hat{p}_j$ and
$\hat{Q}$ $\!=$ $\!\sum_{k=1}^n\hat{q}_k$ are variables of the respective system
sequences. One may think of two trains of light pulses counterpropagating
through a fiber, that may be regionally doped, or [(c), not shown] a ring
resonator allowing repeated interactions.
}
\end{figure}

Interesting open questions we did not address and left for future work are the
possibility of a symbolic solution for quantum Zermelo navigation in a
time-dependent wind $\hat{H}_0(t)$ that generalizes (\ref{Zermelo}). From the
numerical point of view, the gradient-based analyses in Secs.~\ref{sec3} and
\ref{sec4} are thought as illustrations of proof of concept only (not
addressing the question whether a local search of a target unitary starting at
the identity yields a time-optimal navigation to this target unitary). For a
comprehensive investigation it would be interesting to incorporate refined
approaches such as subspace-selective self-adaptive differential evolution
\cite{Zah15}, Pareto front tracking \cite{Tib12}, two-stage hybrid optimization
\cite{Goe15}, or an algebraic construction of the target unitary \cite{Aie14}.
From the experimental point of view, devising a realization of the modulation
(\ref{Zermelo}) along with controlled interactions with an environment that
obeys the desired bath properties is required. Potential candidates include
Bose-Einstein condensates \cite{Nir11}, trapped ions \cite{Bar11}, atoms in an
optical lattice \cite{Yi12}, or arrays of optical cavities as depicted in
Fig.~\ref{fig7}.
%##############################################################################
%##############################################################################
%-----------------------------------------------------------------------------
\begin{acknowledgments}
This work was supported in part by the Austrian Science Fund (FWF) through
project F04012, and by the Templeton World Charity Foundation (TWCF).
\end{acknowledgments}
%-----------------------------------------------------------------------------
%-----------------------------------------------------------------------------
%\bibliography{z}

%-----------------------------------------------------------------------------
%-----------------------------------------------------------------------------
\end{document}